\documentclass[conference]{IEEEtran}
\IEEEoverridecommandlockouts
\usepackage{cite}
\usepackage{amsmath}
\usepackage{multirow}
\usepackage{soul}
\usepackage{hyperref}
\usepackage{comment}
\hypersetup{
    colorlinks=true,
    linkcolor=blue,
    filecolor=magenta,      
    urlcolor=cyan,
}
\usepackage{caption} % DO NOT CHANGE THIS AND DO NOT ADD ANY OPTIONS TO IT
\usepackage{amsmath}
\usepackage{graphicx}
\usepackage{float}
\usepackage{setspace}
\usepackage{algorithm}
\usepackage{booktabs}
\usepackage[noend]{algpseudocode}

\usepackage{balance}
\def\BibTeX{{\rm B\kern-.05em{\sc i\kern-.025em b}\kern-.08em
    T\kern-.1667em\lower.7ex\hbox{E}\kern-.125emX}}
\begin{document}

\title{``Who can help me?'': Knowledge Infused Matching of Support Seekers and Support Providers during COVID-19 on Reddit
}
\author{
    \IEEEauthorblockN{Manas Gaur\IEEEauthorrefmark{1}, 
Kaushik Roy\IEEEauthorrefmark{1}, 
    Aditya Sharma\IEEEauthorrefmark{2}, Biplav Srivastava\IEEEauthorrefmark{1}, Amit Sheth\IEEEauthorrefmark{1}}
    \IEEEauthorblockA{\IEEEauthorrefmark{1} AI Institute, University of South Carolina, USA
    \{mgaur, kaushikr\}@email.sc.edu
    \{biplav.s, amit\}@sc.edu}
    \IEEEauthorblockA{\IEEEauthorrefmark{2} LNMIIT, India
    17ucs011@lnmiit.ac.in}
}

\maketitle

\begin{abstract}
During the ongoing COVID-19 crisis, subreddits on Reddit, such as r/Coronavirus saw a rapid growth in user's requests for help (support seekers - SSs) 
including individuals with varying professions and experiences with diverse perspectives on care (support providers - SPs).  Currently, knowledgeable human moderators match an SS with a user with relevant experience, i.e, an SP on these subreddits. This unscalable process defers timely care.
We present a medical knowledge-infused approach to efficient matching of SS and SPs validated by experts for the users affected by anxiety and depression, in the context of with COVID-19. After matching, each SP to an SS labeled as either supportive, informative, or similar (sharing experiences) using the principles of natural language inference. Evaluation by 21 domain experts indicates the efficacy of incorporated knowledge and shows the efficacy the matching system.
\end{abstract}


\section{Introduction}
\begin{figure}[t]
    \centering
    \includegraphics[width=0.5\textwidth]{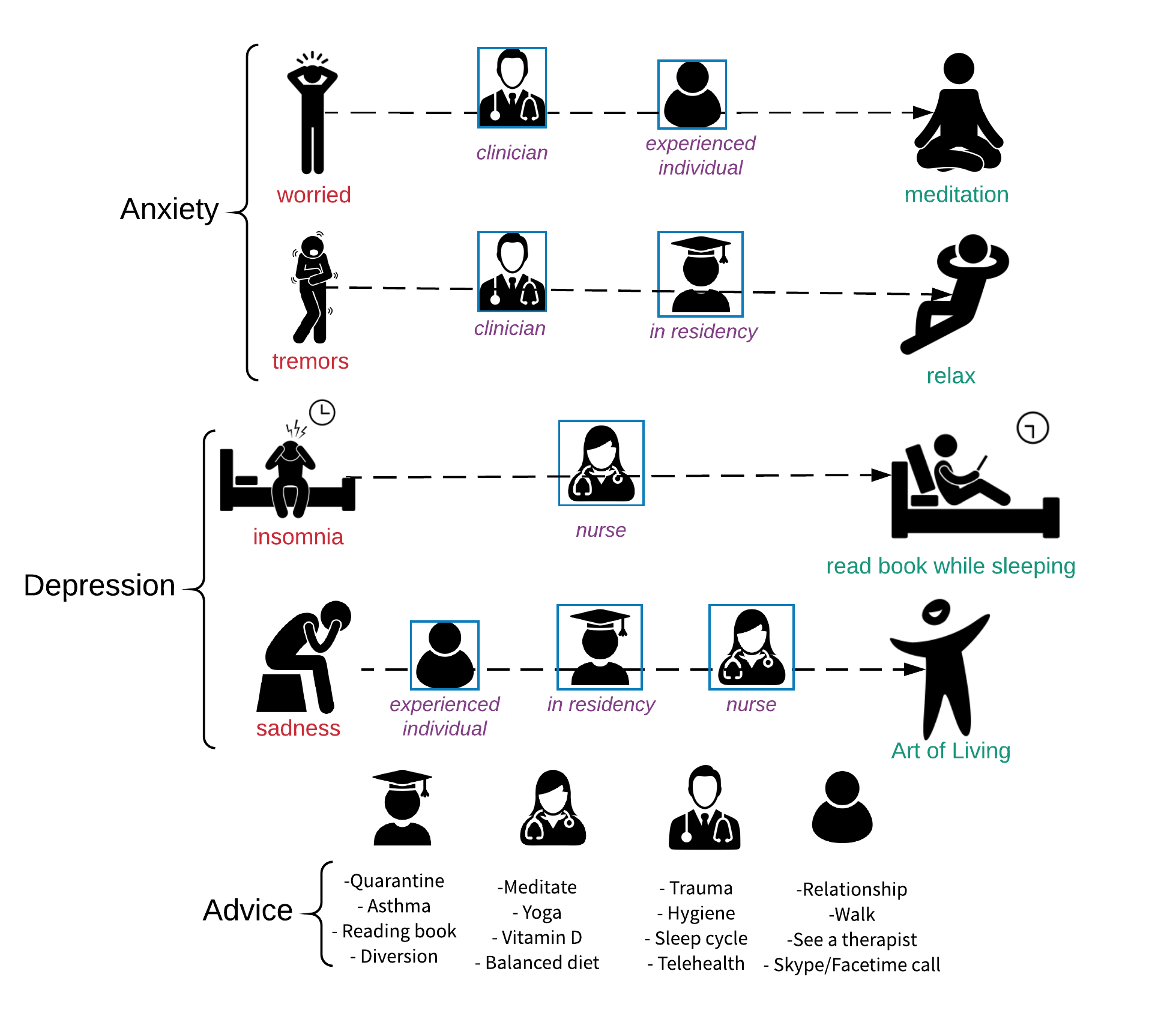}
    \caption{Illustration of a scenario in which users suffering from Anxiety and Depression (SSs) seek help. The moderators endeavor to connect them to the users with the relevant experience (SPs) to address the SS's needs. The moderators may themselves provide advice.}
    \label{fig:figure 1}
%\vspace{-1.6em}
\end{figure}

Within two months of the novel coronavirus pandemic, Reddit’ subreddit \textit{r/Coronavirus’} subscribers jumped from 2,000 to a staggering $\sim$2 Million
with 
%This is best explained by 
``Reddit potentially being the internet’s best support group'' \footnote{\url{http://bit.ly/reddit_support}} \cite{low2020natural}. Subreddits \textit{r/Coronavirus} and \textit{r/covid19\_support} %have become 
are a significant source of information for those coping with Anxiety.
%The success of public health efforts often depends on having a plan to address community concerns. 
%A source like 
Reddit 
%that 
mirrors people’s thoughts and actions,
%can prove 
invaluable in crafting responsive plans. Presently, these subreddits have 64 moderators with diverse backgrounds: (a) academics in genomic science, infectious diseases, virology, and tuberculosis, (b) healthcare workers including nurses, general practitioners, internal medicine specialists, and mental health specialists, and (c) public health personnel and epidemiologists (see 
%illustration of the problem in 
Figure \ref{fig:figure 1}). 
The enormous set of users in these communities has overwhelmed moderators.
%in their ability to provide timely support.
\noindent 
%Compounding the subscriber numbers, 
Worse, users drift in from other mental health %communities
subreddits
(\textit{r/Depression, r/Anxiety, r/Opiates, r/SuicideWatch}) to \textit{r/Coronavirus} %subreddits 
seeking support for their deteriorating mental health conditions due to COVID-19 \cite{alambo2020covid}. Broadly, the conversations on \textit{r/Coronavirus} %consist of %(i) topics related to coping mechanisms such as meditating, Dungeons and Dragons, Ludo King, indoor games, dog walking, online streaming platforms, pleasant conversations, long walks; (ii) topics on various disrupted activities of daily living such as child care/elder care, excessive eating, food issues (e.g., eating bread all day), oscillating sleep cycles, suspended group activities such as choirs, anxieties, frustrations,   helplessness, chronic health conditions such as lung disease, pregnancy, dislocated families, family discord, etc., (iii) 
%discussions s
surround major events %during COVID-19 
such as \{business closure, school closure, lockdown, shelter-in-place, hospitalization\} (also  non-pharmaceutical interventions (NPIs)), and activities of daily living. We use 
%published 
mental health lexicons 
%using 
from the Diagnostic and Statistical Manual of Mental Disorders (DSM-5)\cite{gaur2018let}, depression using Patient Health Questionnaire-9 (PHQ-9)\cite{gaur2021characterization}, anxiety\cite{RHE16}, and information on NPIs to measure the impact of COVID-19 on Reddit users' %psychology. 
mental health.
%distribution of expressions of anxiety and depression and COVID-19 event related topics. This is done to measure the effect of COVID-19 related policy interventions - Non-pharmaceutical Interventions (NPIs) on Reddit users' psychology \cite{seale2020improving}.

%Some 
Users and moderators started the \textit{r/covid19\_support} community to provide emotional support
%and helpful information to the posted problems 
specific to COVID-19. The community received significant traction during February to March 2020, with user counts reaching 20K. One nurse moderator, 
%who was one of the moderators, 
referred concerned users on the \textit{r/Coronavirus} subreddit to \textit{r/covid19\_support} for support on Anxiety and Depression that is virus-related\footnote{\url{http://bit.ly/reddit_nurse_covid19_support}}. However, the task of associating a support seeker (SS) with a potential support provider (SP) within these dynamic communities 
%is unmanageable for 
dwarfs the limited number of moderators.

\noindent \textbf{Can fine-grained knowledge about users improve matching of SS and SPs on r/Coronavirus and r/covid19\_support subreddits?}
\newline
We used data from \textit{r/Coronavirus} and \textit{r/covid19\_support} communities during the first wave of the COVID-19 - when people started experiencing symptoms of mental illness.
%due to NPIs. %various government policies and issues concerning hospitalization. 
To 
%recognize 
detect topics and issues that reference Anxiety and Depression, we used PHQ-9\footnote{\url{http://bit.ly/phq_lex}}, anxiety\footnote{\url{http://bit.ly/anxiety_lex}}, and DSM-5\footnote{\url{https://bit.ly/dsm5_dao_lex}} lexicons. These medical knowledge resources 
%form our representation of medical knowledge used to construct 
support contextual embeddings of SS's posts. %Corroborating our simulated results convergence to high-quality matches, we find that with real Reddit data, \emph{KI} converges to high-quality matches between SS and SPs. We make use of a Siamese Network architecture to approximate the ratings between SS and SPs. %The pipeline followed in this study is illustrated in Figure \ref{fig:architecture}.
%\begin{figure*}[!htbp]
%   \centering
%   \includegraphics{architecture2.png}
%    \caption{Pipeline followed in the study - Proof of concept by simulation experiments, real world experimentation, and quantitative and qualitative evaluation with domain experts (DE1, DE2, DE3).}
%    \label{fig:architecture}
%\end{figure*}
We summarize our key contributions as follows:
\begin{enumerate}
    \item  \textbf{Dataset}: We include the characteristics of 
support seekers and providers, along with the situation.
%, are pivotal in high-quality dataset creation. 
On the crawled 5.3 Million posts and comments from COVID-19-related subreddits, we used psycholinguistic \cite{pennebaker2001linguistic} and pandemic-related knowledge on events to create a working dataset for matching SSs with SPs. We also leverage a general
gold-standard Reddit dataset on supportive and support-seeking behaviors 
%on Reddit to further enrich the quality of the dataset 
(See Section \ref{section: data}).
\item \textbf{Method}: Automated matching of SS with SPs is challenging 
%due to content and psychological aspect which 
because their content generates dissimilar embedding. %Moreover, 
Health-specific markers of SS differ from SP. Hence, %it is essential to use and 
we adapt a model that emphasizes contrastive pairing and allows knowledge-infusion. We use a convolutional siamese network model to match SS and SPs.
\item \textbf{Quantitative evaluation}: 
%We observe that 
Reddit conversations for an SS and SP match resemble Natural Language Inference (NLI) outcomes of entailment, contradiction and neutral. Hence, we evaluate the matches using the NLI paradigm by labeling the SPs as similar(entailment), supportive(contradiction), and informative(neutral) to an SS problem. %mentioned by an SS.
\item \textbf{Qualitative Evaluation}: We evaluated the best performing method qualitatively with a cohort of 21 Domain Experts (DEs).To the best of our knowledge, this is the first study that leverages diverse domain-specific knowledge to match 
online support seekers and providers.
\end{enumerate}

%users seeking help and users providing help online. 
%Moreover, we notice its immediate importance 
In the pandemic context, %where support seeking concerned with \textit{need for resources} can be full-filled by support provided by 
this ability assists the strategic \textit{distribution of limited resources}.
%These users include Medical professionals (includes population health science experts and nurses)(24\%), faculty in science programs (including cognitive science, psychology, life sciences, and physical sciences) (23.7\%), Ph.D. students (33.3\%), and students enrolled in undergraduate science majors in the US (19.0\%). On average, we received a satisfactory rating of 7.25 for depression and 7.04 for anxiety on a scale of 1-10. 

%\textcolor{red}{add layout of document sections}
 %Gkotsis et al., provided a series of psycholinguistic, language, and emotional features to study mental health content shared on social media \cite{gkotsis2016language}. For our problem, we see both a need for such features and the fusion of clinical knowledge. Such knowledge in the form of semantic lexicons helped us in identifying the users with depression and anxiety. The use of psycholinguistic, language and emotional features allowed us to see the correlation between the user behaviors in the context of depression and anxiety.
%After identifying SS and SP users, we noticed feature-level differences between the users in either group. For example, a higher score in “cognitive processes” was seen in SS than SP, showing a decline in the capability of rational thinking due to COVID-19 induced anxiety. We believe associating such an SS user with an SP user who provides support in improving the cognitive process would be beneficial to humanity.

\section{Related Work}
The stigma surrounding mental health 
%and its impact on help-seeking intentions has forced 
encourages patients to 
%anonymously 
seek peer-support anonymously on community-based platforms, such as Reddit or Talk life \cite{pruksachatkun2019moments}.
Powell et al's 12-item general population survey %provided compelling evidence 
demonstrated that individuals with mental illness seek social support through sensitive self-disclosure\cite{powell2006internet}. Andalibi et al. 
%collected evidence of social support and community building to 
categorized user behaviors into well-established classes: supportive and unsupportive \cite{andalibi2017sensitive}. 
%The design strategies for mental health peer-support from 
O'Leary et al. suggested the importance of user expectations, roles, risk, and clinical knowledge for meeting care demands online \cite{o2017design} with peer support. 
%Powell et al., conducted a 12-item general population survey to assess the importance of the internet for individuals seeking information \cite{powell2006internet}. The survey's conclusion was compelling with responses from individuals who had a mental illness and asked for social support through sensitive self-disclosure. Likewise, 
%Cuijpers et al. highlighted the importance of internet-based interventions for mental healthcare by evaluating its outcome on patients compared to those obtained via face-to-face interactions \cite{cuijpers2008internet}.
%Investigative quantitative research from Andalibi et al., collected evidence of social support and community building through visual and textual content on Instagram. The study categorizes user behaviors into well-established classes of supportive and unsupportive behaviors \cite{andalibi2017sensitive}. 
%Existing mental health peer support technology is limited to diagnosis, and role identification does not match peers, and therefore cannot proactively mitigate risk through intervention and treatment information.
%O’Leary et al., designed strategies for mental health peer-support \cite{o2017design}. The study highlights the need for technology that accommodates user expectations, roles, risk and adequately compensates for the lack of professionals in meeting care demands. 
%For instance, 
Gillani et al., showed that a simple reframing of irrational thoughts through support provider perspectives could bring positive cognitive change to those suffering from mental stress \cite{gillani2018me}.
However, prior research did not employ contextualization and abstraction of social media content, a requirement for capturing health-specific markers of an SS user and match with an SP, who can provide relevant help \cite{gaur2021semantics}.
%Thus 
We build and improve upon %some 
related research \cite{purohit2014identifying, yang2019seekers} on SS and SP role identification by forming dynamic support groups using psycholinguistic and clinical knowledge braided with deep learning.

\section{Exploratory Data Analysis}{\label{section: data}}
%We perform exploratory data analysis on the posts to determine the right features to use for the real world matching model, as well as determine the simulation settings 
%To measure the surge in anxiety and depression related concerns correlated with COVID-19, w
We collected nearly 5.3 Million posts and comments from nearly 450K users in \textit{r/Coronavirus} subreddit and 48K posts/comments from 6.9K users in \textit{r/covid19\_support} subreddit (see Table \ref{tab:data-stats}).

%To identify surges of mental health Reddit posts due to COVID-19, our study collected   

\begin{table}[!htbp]
\begin{tabular}{p{3.5cm}|p{4.5cm}}
\toprule[1.5pt]
\multirow{2}{*}{Timeframe} & \textit{r/covid19\_support}=February 22 to May 31, 2020 \\
& \textit{r/Coronavirus} = January 13 to May 13, 2020 \\\midrule[1pt]
\#Posts in subreddits & \textit{r/Coronavirus}= 5.3M, \textit{r/covid19\_support}=48K \\\midrule[1pt]
\#Users in subreddits & \textit{r/Coronavirus}=523K, \textit{r/covid19\_support}=7K \\\midrule[1pt]
\#Users per mental health condition & Anxiety=8K and Anxiety-SS = 6.8K, Depression= 4K and Depression-SS=2.7K   \\\midrule[1pt]
\#Posts per COVID-19-related events & Business Closure=14.7K, School Closure=13K, Lockdown= 27.6K, shelter-in-place=8K, Hospitalization=1K
 \\
\bottomrule[1.5pt]
\end{tabular}
\caption{Reddit 
%Data 
statistics for posts and users. The time difference between r/Coronavirus and r/covid19\_support subreddits is due to the %late 
delayed opening of r/covid19\_support 
%based out of the for 
for help-seeking 
%intentions of 
users in the r/Coronavirus subreddit.}
\label{tab:data-stats}
%\vspace{-2em}
\end{table}

\begin{figure}[!h]
\centering
\includegraphics[width=0.5\textwidth]{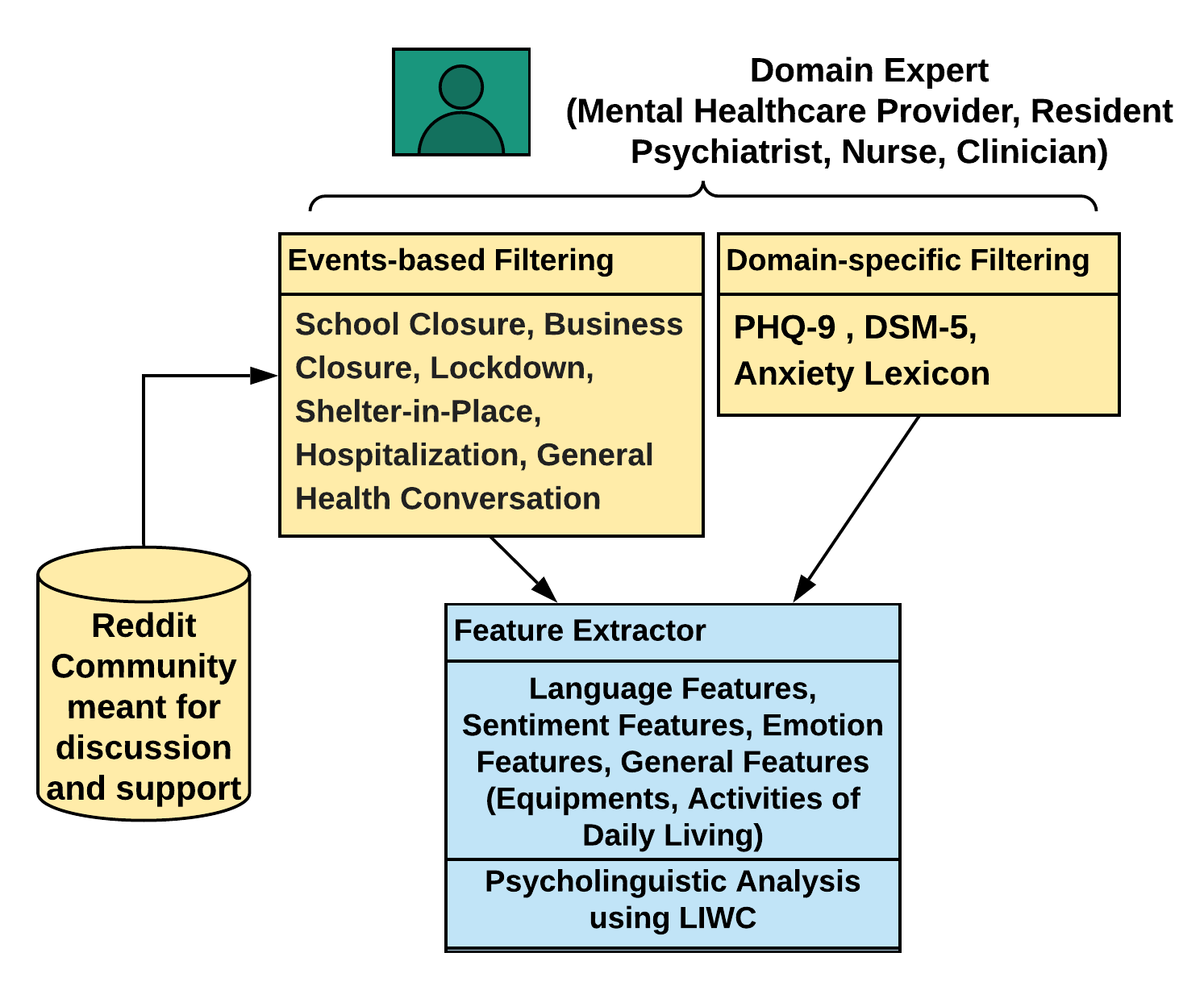}
\caption{Feature extraction pipeline distilling noisy Reddit posts by significant COVID-19-events, and clinical terms on anxiety and depression; %following which, 
subsequently we generate language, sentiment, emotion, ADLs, and psycholinguistic feature vectors along with content to representation SS and SPs}
\label{fig: EDA}
\end{figure}

The DEs involved in this study noted that 
%examined the data and made two inferences: 
(1) Typical comments and posts in \textit{r/covid19\_support} subreddit are either supportive/helpful, and (2) There is a fair distribution of problem-focused and informative comments/posts in \textit{r/Coronavirus} subreddit, but rarely a supportive post/comment. Thus, we 
%created the dataset by 
merged the post/comments from \textit{r/Coronavirus} and \textit{r/covid19\_support} subreddits. From this large-scale data, we constructed a subset of users with their posts/comments specific to mental health (Anxiety and Depression) and events in COVID-19 (e.g., business closure, school closure). Figure \ref{fig: EDA} shows the entire exploratory data analysis pipeline
%We organize the process of constructing the experimental data as follows (see Figure \ref{fig:figure 3}):
%\begin{figure*}[!t]
%    \centering
%    \includegraphics[width=\textwidth]{aamas21-formatting-instructions/figure_3.png}
%    \caption{Design of the matching system for a community-based platform for user-safety in the context of Mental Healthcare with feature descriptions, soft clustering and NLI-based user role identification}
%    \label{fig:figure 3}
%\end{figure*}

\textbf{Event-based Filtering:} 
School closure, business closure, lockdown, shelter-in-place, and hospitalization, and general COVID-19 conversation during COVID-19 likely had a global psychological impact (see Table \ref{tab:dist}).
We filtered posts and comments in the raw data through soft matches with COVID-19-related events  using embedding with a cosine similarity threshold over 0.8. %Following this filtering, 
Subsequently we
%we further 
filtered posts and comments 
for 
%that expressed 
concerns regarding Anxiety and Depression through SS or SP behavior.  

\textbf{Identify Users with Anxiety and/or Depression:}
We used an anxiety lexicon and a negation detection method to tag complex posts or comments in the dataset correctly. The lexicon captured varied 
expressions 
%mentions 
of anxiety, 
%including terms like 
such as anxiety, anxiousness, anxious, agita, agitation, prozac, sweating, and panic attacks.
For instance, the following post: 
``Then others that insisted that what I have is depression even though \textit{manic episodes} \underline{aren't} \textit{characteristic of depression}. I dread having to retread all this again because the clinic where I get my mental health addressed is closing down due to loss in business caused by the pandemic'' was tagged as "Business Closure with Anxiety" as italicized phrases
%are present 
appear in the semantic lexicon. 
Negation detection identified \underline{``aren't,"} as the negation, making this sentence as ``not depression," and tagging it as ``anxiety." 
Table \ref{tab:dist} notes that hospitalization and lockdown events caused relatively higher anxiety levels in individuals compared to business closure, school closure, and shelter-in-place. %An analysis of general COVID-19 conversations in the context of anxiety showed an increased unemployment rate, scarcity of PPEs, and continence management (presumably due to lack of movement and toilet paper hoarding) as factors for increased anxiety levels, causing severe panic attacks.
\begin{table}[]
    \centering
    \begin{tabular}{c|c|c|c|c|c}
    \toprule[1.5pt]
     \textbf{Psy}/\textbf{\textit{C}}/\textit{Ev}  & D & A & \textbf{Psy}/\textbf{\textit{C}}/\textit{Ev} & D & A  \\
     \midrule[1pt]
     \textbf{Emotional} & 0.39 & 0.5 & \textbf{Social} & 0.24 & 0.6 \\
     \textbf{Biological} & 0.23 & 0.48 & \textbf{Cognitive} & 0.25 & 0.53 \\
     \textbf{Future} & 0.001 & 0.23 & \textbf{Modals} & 0.31 & 0.64 \\
     \textbf{\textit{Inst ADLs}} & 0.21 & 0.4 & \textbf{\textit{Basic ADLs}} & 0.18 & 0.3 \\
     \textbf{\textit{Equipment}} & 0.22 & 0.33 & \textit{Sch Closure} & 0.21 & 0.23 \\
     \textit{Bus Closure} & 0.2 & 0.31 & \textit{Lockdown} & 0.22 & 0.29 \\
     \textit{Shelter} & 0.2 & 0.37 & \textit{Hospital} & 0.24 & 0.4 \\
     \bottomrule[1.5pt]
    \end{tabular}
   \caption{Correlations between depression (D) or anxiety (A) posts and \textbf{psycholinguistic features (Psy)}, \textit{\textbf{COVID-19 related terms (C)}} (bold + italics) and \textit{COVID-19 related events (Ev)}}
   %Distribution of \textbf{psycholinguistic features (Psy)}, \textit{\textbf{COVID-19 related terms (C)}} (bold + italics) and \textit{COVID-19 related events (Ev)} in people who mention depression (D) or anxiety (A). Scores represent correlation between D/A and Psy/C/Ev features.}
    %\caption{\footnotesize }. 
    \label{tab:dist}
    %\vspace{-1.4em}
\end{table}
%\begin{figure}[!h]
%    \centering
 %   \includegraphics[width=0.5\textwidth]{figure_4.png}
%    \caption{\footnotesize The distribution of users mentioning depression and anxiety within the context of various major events during COVID-19. Scores represent correlation between events and mental health conditions.}
 %   \label{fig:figure 4}
%\end{figure}
For example, posts like ``\textit{Need help. Have a friend who lives alone who is now suicidal from the isolation and anxiety, and already had depression. I've asked her to come to my house for the shelter in place, but she doesn't want to}" were tagged as "Shelter-in-place with Depression and Anxiety." %Furthermore, a SS user's post: "\textit{I am freaking out. I am so scared of getting the virus I \textit{can't sleep}, 
%\textit{can't stop thinking} about myself in the ICU, or dying in my home \textit{gasping for air}. I tried \textit{meditation}, \textit{talking with family}, with a psychologist, but nothing seems to help. Does anyone have suggestions?}" was tagged as "General COVID conversation and Depression." %In the elevated levels of adverse mental health conditions during COVID-19, Figure 3 reports more users expressing depression than anxiety in the concerned subreddits.Health-care workers and the elderly are among the worst affected due to the constant suffering they witness. Unemployment from closures also caused depression in isolated individuals, unable to meet their needs. Anxiety was comparably heightened, e.g., dealing with the fear of death in addition to the anxiety from uncertainty in the resumption of their livelihood
%\footnote{\url{https://www.washingtonpost.com/health/2020/06/07/health-care-workers-coronavirus-burnout/?arc404=true}}.

\textbf{Feature Extraction:}
We discuss two types of features 
%in our study 
that characterize SSs and SPs:
\paragraph{Psycholinguistic features}
Given 
%Once 
tagged users with the most relevant mental health condition (anxiety or depression), 
%are tagged, 
we extracted 
%psycholinguistic 
lexical features describing 
%various topics of 
concerns surrounding 
cognitive, emotional (positive emotions, negative emotions, affect, anger, sad), biological (sexual, body, ingest, health), Focus Future, and social processes (Social, Family, Friend). LIWC\footnote{\url{https://liwc.wpengine.com/}} (linguistic Inquiry and Word Count) provides a comprehensive categorized dictionary of words that people use words in their daily lives. This provides rich information about their beliefs, fears, thinking patterns, social relationships, and personalities in crisis \cite{pennebaker2001linguistic}. 
\paragraph{COVID-19 features} Modals (epistemic and dynamic), Instrumental Activities of Daily Livings ( Inst ADLs), Basic ADLs, and Equipment are additional features in conversations specific to COVID-19. Instrumental ADLs include ``moving within the community,'' ``preparing meals,'' ``using telephone for communication,'' ``cleaning and maintaining the house,'' ``taking prescribed medication,'' and ``managing money''. Basic ADLs comprise ``personal hygiene,'' ``toilet hygiene,'' ``functional mobility (e.g., getting in and out of bed), and self-feeding.
LabMT, a Mechanical Turk emotion assessment tool, provided a real-valued score for emotional features
%The 
%A real-valued score of emotional features was calculated using LabMT, a Mechanical Turk language assessment for emotions.
%by Mechanical Turk 
\footnote{\url{https://trinker.github.io/qdapDictionaries/labMT.html}}. 
%From prior research on language analysis of Reddit, we considered the following features as ``language features'': maxHeight of the posts (length of the dependency parse tree), noun chunks, maximum verb phrase length, subordinate conjunctions, pronouns, and first-person pronouns \cite{gaur2019knowledge}.

\textbf{General Findings:} 
%On the filtered data created from the sequential processing of event-based filtering, domain-specific filtering, and feature extraction, w
We note the following (see Table \ref{tab:dist}). 
%\begin{figure}
%    \centering
 %   \includegraphics[width=0.5\textwidth]{figure_5.png}
 %   \caption{\footnotesize We record the distribution of various psycho-linguistic and COVID-19-related behavioral features (X-axis) impacted by COVID-19, against people that mention depression or anxiety (Y-axis). Scores represent correlation between Depression/Anxiety and psycholinguistic/covid-related features.}
%    \label{fig:figure 5}
%\end{figure}
Users who expressed depression 
%in our collected dataset 
were affected by the disruption in social processes, epistemic modality (\textit{described as pandemic uncertainty with words like may or might} %in a sentence)
and dynamic modality (\textit{described as possible or necessary conditions expressed by words like can or will in a sentence}) (see Table \ref{tab:dist}). COVID-19 has impacted ADLs. The effect 
%has been 
is 
%more significant 
greater for users with depression compared to users with anxiety. Depressed individuals used words that reflected concerns on their 
%future 
post-COVID-19 futures, which we capture by mapping them to term under \textit{Focus Future} category in LIWC \footnote{\url{https://www.helpguide.org/articles/anxiety/dealing-with-uncertainty.htm}}. The %intensity 
association of such words was higher in depressed individuals compared to users with anxiety. Questions or concerns such as adjustment with traumatic events, how to recover from anorexia, having difficulty in thinking, and poor expression of thoughts scored higher on the cognitive process %psycholinguistic feature 
category for users with depression than with anxiety. The frequent usage of phrases similar to “disruption in sleep”, “low body vitals”, and “sexual words” suggested an impact on the biological processes %psycholinguistic 
category during COVID-19. This 
%was seen to be
appeared equally prominent in users with depression and anxiety.
Further, conversations mentioning isolation, social distancing, and loneliness mainly came from users with depression compared to those with anxiety. Emotional processes of both users with depression and anxiety were equally affected, showing significantly high scores in anger, frustration, and sadness. Situations like “unable to get prescribed drugs” lead to more anxiety concerns than depression concerns. \textcolor{black}{We performed statistical significance testing with Bonferroni-correction (p-value$<$0.05), showing that features in table \ref{tab:dist}) are discriminative for anxiety-depression.}% T-test was performed on the results reported in Table 5:-(a)-Non-rejection of null hypothesis across four professions, signifying a similarity in the selection of SPs among individuals with same professions (95\%-Confidence intervals, values in Table-5)} %\vspace{-0.3cm}

\textbf{SS and SP classification:} %\begin{figure}
%    \centering
%    \includegraphics[width=85mm, height=65mm]{roc.png}
%    \caption{The ROC plots compare the capability of the approaches in detecting supportive/non-supportive users in the labeled dataset on suicide risk severity. We notice that GPT-2 with logistic regression outperformed other methods.  These are mean scores from 5-fold cross validation}
%    \label{fig:roc}
%    \vspace{-1em}
%\end{figure}
To classify users as either SS or SP, we used a labeled dataset from a recent study on the assessment of suicide risk.
%severity research. 
The reported Krippendorff agreement was 0.79, sufficient to be a gold standard \footnote{\url{https://zenodo.org/record/2667859\#.X_kTIi2z1hE}} \cite{gaur2019knowledge, gaur2021characterization}.
%The dataset consists of nearly 30K comments with an average of 16 sentences per comment from a filtered set of 500 users extracted from r/SuicideWatch and other mental health subreddits (e.g. r/Anxiety, r/Depression, r/Bipolar, r/StopSelfHarm, etc.) using a clinical questionnaire on suicide risk severity \cite{posner2011columbia}. In this dataset, a large fraction of users mentioned borderline personality disorder, depression, addiction, and anxiety. The dataset contains five labels: supportive, indication, ideation, behavior, and attempt. Before training a classifier for our study, we reduce the number of labels to two: supportive (or SP) (108 users) and non-supportive (SS = indication, ideation, behavior, and attempt) (392 users) as we do not require the other labels.
Over the annotated dataset, we experimented with various 
%approaches involving a 
pipelines, generating contextualized embedding followed by a classifier to predict SP or not SP. For generating embeddings of the posts/comments, we used: ELMo \cite{peters2018deep}, Universal Sentence Encoder (USE), and GPT2 Embeddings \cite{cer2018universal} and for classification, we used: Logistic Regression (LR), and Support Vector Machines.  GPT2+LR is 
%found to be 
the best performing 
%procedure for classification in the 
classifier for this dataset. We chose these representation learning models as they are State-of-the-Art and can generate context-aware embeddings. Furthermore, we do not use BERT as it requires input chunking 
%of the input 
due to its max token length of 512 \cite{devlin2018bert}. We used the GPT2+LR model to generate probabilities of a user being SS in our dataset. We do not label SPs as we know the ground truth SPs annotated by DEs to be 108. We labeled approximately $9.5$K ($6.8$K Anxiety and $2.7$K Depression, Table \ref{tab:data-stats}) users as $SS$ and found that the SS:SP ratio of $\sim 10K$:$100$ reflects the true distribution on Reddit.
%\vspace{-0.1em}
\begin{table}[ht]
\begin{tabular}{p{3.0cm}p{0.5cm}p{0.5cm}p{0.5cm}p{0.5cm}p{0.5cm}p{0.5cm}}
\toprule[1.5pt]
\multirow{2}{*}{Methods} & \multicolumn{2}{c}{Precision} & \multicolumn{2}{c}{Recall} & \multicolumn{2}{c}{F1-Score} \\
 \\ \cmidrule{2-7}
 & SS & SP & SS & SP & SS & SP \\ \midrule[1pt]
 Without KI & 0.79 & 0.48 & 0.79 & 0.48 & 0.79 & 0.48 \\
 (Content) &&&&&& \\ \midrule
 With KI & 0.88 & 0.68 & 0.62 & \textbf{0.90} & 0.73 & 0.77\\
 (Psy, $P_{SS,SP}$)&&&&&& \\ \midrule
 With KI & 0.72 & 0.88 & \textbf{0.94} & 0.55 & 0.82 & 0.68 \\
 (Psy, &&&&&& \\
 $P_{SS,SP}$,  COVID-19)&&&&&& \\ \midrule
 With KI & \textbf{0.89} & \textbf{0.89} & 0.92 & 0.86 & \textbf{0.90} & \textbf{0.87} \\
 (Content, Psy, &&&&&& \\
 $P_{SS,SP}$,  COVID-19)&&&&&& \\ 
 
\bottomrule[1.5pt]
\end{tabular}
\caption{
%Performance of without KI and with KI strategies. 
An Ablation study with KI with three possible configurations: Psy: Psycholinguistic Knowledge, $P_{SS,SP}$ = Pr(X=\{SS,SP\}). 
%We see that 
With Psycholinguistic knowledge, the recall on SPs increases. 
%significantly. On the other hand, 
However, SS recall rises with the addition of COVID-19 knowledge.
%SS recall rises. 
%When we 
Combining the two
%, there is a significant improvement in 
improves performance 
%across the board. 
overall. 
%(relative improvements in the ablation study are bold-faced.)
Ablation study improvements are bold-faced.}
\label{tab: table 100}
\end{table}
\vspace{-0.5em}
\section{ Knowledge Infused Match Prediction}
\label{sec:M}
\begin{table*}[!htbp]
\begin{tabular}{p{0.48\textwidth}p{0.48\textwidth}} 
\toprule[1.5pt]
\textbf{Problems} & \textbf{Responses} \\
\midrule[1pt]
\multirow{3}{22em}{I am not sleeping much anymore. \textbf{Anxiety} is pretty high for the stability of the world and the future of trust. Probably need to take up drinking or something...
} & \textit{Supportive}: Giving up is in your control. Exercise can be lots of different things and a way to help anxiety.
\\\cmidrule{2-2}
& \textit{Informative}: Anxiety is quite inducing. A good time to learn relaxation techniques \\\cmidrule{2-2} 
& \textit{Similar}: I hear you. Myself and other friends with kids are going through similar anxiety right now [...]  \\ \midrule[1pt]
\multirow{2}{25em}{Married with a supportive husband but my serious health issues including \textbf{depression and PTSD} has made me feel as if I am losing everything [...]} & \textit{Supportive}:Tough position [...] kind of relate [...] don't think marriage is dying [...] advice would be to meditate, prioritize, and act patiently. Be positive [...]
\\ \cmidrule{2-2}
& \textit{Similar}: My MDD is affecting my married life. I am an outdoor enthusiast and so is my husband. My health concerns keep pulling him down [...]
wanted him to let me go. \\ 
\bottomrule[1pt]
\end{tabular}
\caption{Examples of SS problems and SP responses categorized as \textit{Supportive}, \textit{Similar} or \textit{Informative} by the NLI model: RoBERTa. It can be seen, how \textit{Supportive} posts contradict the view in the SS's post (``\textit{My point is anxiety is worse than death, Do not let your mind slip}''), \textit{Similar} posts are entailed in the SS's post (``\textit{I hear you, I am on the same boat})'', and \textit{Informative} posts are neutral, specifying guidelines and coping mechanisms.}
\label{tab: table 4}
\end{table*}

In the proposed approach, Knowledge-infused Match (KI-Match), we train a Convolutional Siamese Network Model with contrastive loss to predict matches, computed using the knowledge features and the GPT-2 encoding of the posts, using low dimensional representations of the users
for details on the architecture please see \cite{vilhagra2020textcsn} %Figure 1). The loss function used is
           \[ CoSim(SS,SP) - CoSim(SS,\overline{SP}) + \alpha  \leq 0 \]
Where SS is the support seeker post, SP is a relevant Support Provider, and $\overline{SP}$ is a non-relevant support provider. \textit{CoSim} is the cosine similarity between the two data points, and \( \alpha \) is the margin.
Specifically, we pass a concatenated vector of the GPT-2 encoded post vector and the features extracted, namely the pyscholinguistic features (Psy), and COVID-19 features into the siamese network blocks to construct low dimensional representations. Each Siamese Network block is a Convolutional Neural Network, and the supervision signal is binary, i.e. Label 1 denotes a good match and 0 a bad one. %and as calculated as described next. For a pair of posts from an SS and an SP, let $P_{SS}$ and $P_{SP}$ represent the KI vector representations of the post.

\section{Results and Analysis}
%\begin{table}[!h]
%\small
%\begin{tabular}{p{1.8cm}|p{0.5cm}p{0.5cm}p{0.5cm}p{0.5cm}p{0.5cm}p{0.5cm}p{0.5cm}}
%\toprule[1.5pt]
%\textbf{Disorders} & \textbf{SS} & \textbf{S} & \textbf{I} & %\textbf{Ex} & \textbf{S \& I} & \textbf{I \& Ex} & \textbf{S \& I \& Ex}  \\ 
% \midrule[1pt] %\hline
% Anxiety (65 clusters) & 649 & 178 & 1959 & 2 & 1266 & 10 & 18\\  
% Depression (53 clusters) & 399 & 126 & 1379 & 7 & 895 & 2 & 8\\ 
% \bottomrule[1.5pt]
%\end{tabular}
%\caption{Distribution of SP vs SS after the clustering process as identified by NLI paradigm. SS: Support Seeker, S: Support by prevention, I: Support by information, Ex: Support by experience, S \& I, I \& Ex: User with both roles, S \& I \& Ex: Users with all three roles. Note: SP = S + I.}
%\label{tab: table 3}
%\end{table}
%\vspace{-1cm}

\textbf{Quantitative Evaluation:} After hyper-parameter tuning, we compared the best performing Siamese Network models using precision, average recall, and f1-score (See Table \ref{tab: table 100}). 
%The first model is
GPT-2 clustering without KI \cite{kieuvongngam2020automatic} is above the different configurations of knowledge models.  %It is evident that 
KI methods with different forms of knowledge in the form of psycholinguistic features, COVID-19 features, and probability of support provider/seeker improves the accuracy of match prediction. 
%the right match. %TSNE clusters in Figure \ref{fig:TSNE} further illustrate visualization of match quality for unsupervised clustering and the models in Table \ref{tab: table 100}. Here it can be seen, the cluster quality steadily improves with the addition of more relevant/contextual knowledge. %\textcolor{red}{Quantitative Inspection - Expand}
%\begin{figure*}[t]
 %   \centering
 %   \includegraphics[width=\textwidth]{tsne_plots_fig_8.png}
 %   \caption{TSNE visualization of SS-SP embeddings clustering without knowledge (A), with psycholinguistic knowledge (B), with psycholinguistic and COVID-19 knowledge (C), and with psycholinguistic, COVID-19, and content knowledge (D). A nuanced distribution of SSs and SPs in (D), shows a substantial success of the proposed matching approach. We remark that representation of SS and SP produced using the contrastive loss function in the siamese networks allows the clustering of users with roles SS and SP.  }
%    \label{fig:TSNE}
%\end{figure*}

\begin{table}[!ht]
\centering
\begin{tabular}{ccccc}
\toprule[1.5pt]
\textbf{Cohort} & \textbf{C1}
 & \textbf{C2} & \textbf{C3} & \textbf{C4}
 \\ 
 \midrule[1pt]  %\hline
 Faculty (5) & 2.96 & 7.16 & 2.24 & 6.84 \\  
 MPs (5) & 3.08 & 8.4 & 2.8 & 7.92 \\   
 Ph.D. (7) & 2.26 & 6.45 & 2.0 & 6.85 \\
 UG (4) & 2.65 & 6.15 & 2.55 & 7.40 \\
 \bottomrule[1.5pt]
\end{tabular}
\caption{Mean \#SPs selected by experts from 4 recommendations per SS having anxiety problems (\textbf{C1}). Their confidence (on scale 1 to 10) in selection is reported in \textbf{C2}. Likewise for Depression, mean \#SPs selected by experts from 4 recommendations per SS and their confidence is reported in \textbf{C3} and \textbf{C4} respectively. MPs: Medical Professionals.
%Due to the high confidence of medical professionals, the effectiveness of incorporating medical knowledge in the matching system is evident. Furthermore, relatively lower confidence in faculties and their Ph.D. students offers a different perspective requiring encoding of knowledge on human behaviors in cognitive science and life science.
}
\label{tab: table 5}
\vspace{-1em}
\end{table}

\textbf{Qualitative Evaluation:}
%In the cohort-based assessment of the proposed method, we noticed relatively high confidence scores among different members, which account for the differences in their perspectives. High confidence provides strong qualitative support for medical knowledge preservation using our method. The implicit reference to known mental health conditions is recognized through these lexicons (medical knowledge preservation), which resulted in the precise labeling of users as suffering from either anxiety or depression. %\noindent\textit{Comparison to Natural Language Inference (NLI):}
Previous work 
%has covered 
improved the learned representations of neural network models using NLI methods. 
Instead, we 
%differ by using 
use NLI in labeling the SPs 
%that are 
matched to an SS. %Therefore we evaluate our learned representations using the NLI paradigm. 
 In general, NLI compares a premise statement with a hypothesis statement \cite{sinha2020learning,poliak2018collecting,conneau2017supervised,dusek-kasner-2020-evaluating}.
NLI outputs labels on the hypothesis as, \emph{entailment} - entailed from the premise, \emph{contradiction} - contradicting the premise  and \emph{neutral} - neutral to the premise. We consider the premise as the SS post and the hypothesis as the SP post. 
%We expect that the 
\textit{Supportive} posts from SP users should contradict the SS posts, \textit{Similar} SP users will be entailed from the SS posts, and \textit{Informative} SP users will be neutral to the SS's post. Using this labeling scheme and a pre-trained and fine-tuned RoBERTa transformer model, we determined the NLI labels for SPs
(see 
%A sample of labeling is presented in 
Table \ref{tab: table 4}). %We generated 65 clusters (Anxiety) and 53 clusters (Depression) of SS and SP based on the variation in users' information. Given the  soft clustering method, there is a significant difference in the overlap between SSs and SPs across multiple clusters due to the diversity in SP responses to SSs. In the NLI experiment, we first selected 18\% of total Anxiety and Depression users clustered as SS and SP. The selection criteria were heuristically defined by the high value of the \textit{first-person pronoun ratio} and large gap between the predicted probabilities of a user being SS or SP (SS$>>$SP) by the classifier described in Section 3\ref{3e}. Specifically, for NLI, such a heuristic was constructed to extract those users who self-report their conditions and seek support. This resulted in 649 SS users with anxiety and 399 SS users with depression. Table \ref{tab: table 3} provides the distribution of users labeled as Supportive ( label provided by NLI model: Contradiction), Informative (label provided by NLI model: Neutral), and Similar Problem/Experience (label provided by NLI model: Entailment).
%In Table \ref{tab: table 3}, we see 3 SP users per problem user for anxiety and  4 SP users per problem user for depression (Table 4: (S + I)/P). In contrast, we notice that the total of supportive and informative users per problem user is much greater in anxiety than discussion. Inspection of the posts by the cohort revealed that the nature of helpful responses to anxiety is more informative due to the ability to pin down anxiety to a particular set of parameters such as heightened fear and other related symptoms (see Table \ref{tab: table 4}). For depression, the manifestation of the disease in individuals occurs with much greater variability. Hence, the nature of helpful responses is long story-based rather than prescriptive suggestions. 
Further, a cohort-based assessment of matched SS and SPs (also labeled) from 21 domain experts showed 75\% relevance agreement 
(they consider 3 out of 4 recommended SP users as relevant to an SS user) 
with the KI-Match prediction model. Each expert was asked: 
%two questions as they see the matched SS-SPs: 
(1) How many \textbf{Users} can provide support/help/information of use to the \textbf{Problem User}. (2) What is your confidence in the response? on a random set of 21 SSs expressing anxiety, 21 SSs expressing depression and each having 4 recommended SPs per SS.
As 
%seen in 
Table \ref{tab: table 5} shows, Faculty, Ph.D. students and undergraduates (UG) gave a confidence score of 7 out of 10, whereas medical professionals gave a confidence score of 8/10 in predicted matches. 
Thus, 
some variance remains in the shallow infusion of knowledge occurring through psycholinguistic and language-oriented features.
%may not fully capture their psychological perspectives.

%On average, they consider 2 out of 4 (50\%) recommended users as relevant to an SS user. 
%Academics research the pros/fallacies in the current medical knowledge bases and form their perspectives that perhaps accounts for their differing confidence. 
%We believe a deeper infusion of knowledge through abstraction using categories described by faculties of life science, cognitive science, and psychology would improve SS and SP users' relevance-based matching. We see that UGs generally agree with the recommendations, and this is encouraging as they are the primary users of the Reddit platforms.
%\vspace{-0.8em}
%\section{Implications}
%\input{discussion}
%\vspace{0.5em}
\section{Conclusions and Future Directions}
%The results of this study are preliminary and exploratory as it achieved an above-average response from the cohort. 
%Some significant 
Limitations of this research include: (1) The comparison of the quantitative aspects of SS-SP matching to NLI definitions is analogous to common sense understanding. However, surveying members who occupy such roles on Reddit would allow us to compare with user perception. (2) Infusing clinical and psycho-social knowledge could be improved through attention-based methods. (3) Mental health-specific questionnaires and expert-QAs could be used to match SS with SPs with explanations. 
These require 
%We would perform all this under 
behavioral therapists' supervision.

%\noindent \emph{Reproducibility and Deliverables:}
%Results described in this study through our code and the datasets will be made public post-acceptance. The interface design of the 
%tool used for 
%evaluation tool is available \href{http://bit.ly/SS_SP_tool}{here}.
%\vspace{-1em}

\section*{Acknowledgement}
We acknowledge partial support from National Institutes of Health (NIH) award: MH105384-01A1: ``Modeling Social Behavior for Health-care Utilization in Depression''. Any opinions, conclusions or recommendations expressed in this material are those of the authors and do not necessarily reflect the views of the NIH. 
%\section{Conclusion}
%\section{Challenges and Future Work}
\bibliographystyle{IEEEtran} 
\bibliography{references}

\end{document}